\documentclass[prb,twocolumn,showpacs,superscriptaddress,preprintnumbers,amssymb]{revtex4}
\usepackage{graphicx}
\usepackage{dcolumn}
\usepackage{bm}
\usepackage{wasysym}

\newcommand{\beq}{\begin{equation}}
\newcommand{\eeq}{\end{equation}}
\newcommand{\beqn}{\begin{eqnarray}}
\newcommand{\eeqn}{\end{eqnarray}}

\begin{document}
\title{Phase transitions in coupled two dimensional XY systems with spatial anisotropy}
\author{Cenke Xu}
\affiliation{Department of Physics, University of California,
Berkeley, CA 94720}
\date{\today}
\begin{abstract}

We study phase transitions of coupled two dimensional XY systems
with spatial anisotropy and $U(1) \times \mathbb{Z}_2$ symmetry,
motivated by spinless bosonic atoms trapped in square optical
lattice on the metastable first excited $p-$level orbitals with
anisotropic hopping amplitudes. The phase transitions of the
system are generally split into an Ising transition and an XY
transition, but the sequence and the nature of the transitions
depend on the ratio between the anisotropic couplings. In the
isotropic limit the XY variables are expected to be disordered
before the Ising variables when thermal or quantum fluctuations
are turned on gradually. In the anisotropic limit with zero
perpendicular hoppings, the finite temperature transition is a
Kosterlitz-Thouless transition driven by proliferation of hybrid
half vortices, and the zero temperature quantum phase transition
is split into a bond order transition and a 3D XY transition,
which can be driven by the condensation of either single vortices
or half vortices. After the condensation of half vortices the
resultant state is a Mott Insulator of paired bosons. A small
perpendicular hopping $J_b$ leads to a 2D Ising transition at low
temperature and a 2+1d quantum Ising transition with a small
charging energy at zero temperature. Global phase diagrams for
both classical and quantum phase transitions are drawn. The
analytical results obtained in this work are expected to be
checked both numerically and experimentally.

\end{abstract}
 \maketitle

Ever since the observation of the quantum phase transition between
the Mott Insulator and the superfluid \cite{greiner2002}, the
atomic systems trapped in optical lattices formed by laser beams
have attracted a lot of attention. The potential formed by laser
beams can usually be approximated by the harmonic potential. The
single particle ground state wavefunction within each well of the
optical lattice has an approximate $s-$wave symmetry. The first
excited states are three fold degenerate $p-$wave states. As
discussed in reference \cite{girvin2005} and confirmed by recent
experiment \cite{bloch2007}, although the $p-$wave states are not
the ground states, the life time for these states can be much
longer than the average tunnelling time between neighboring
optical lattice sites. If all of the atoms are pumped from the
ground state to the $p-$wave states, they will establish a
metastable equilibrium which can survive for considerable time. In
previous studies, various phases have been identified for this
system, including a superfluid phase with one dimensional
$\mathbb{Z}_2$ gauge symmetry \cite{girvin2005,wu2006a} on square
and cubic optical lattices, a novel stripe phase on triangular
lattice \cite{wu2006}, a bond algebraic liquid phase \cite{xu2007}
and a crystalline orbital order on honeycomb lattice
\cite{wu2007}. Condensate of $p$-wave paired atoms has also been
considered \cite{kuklov2006}. Very recently, the long-life
$p-$wave excited states have been realized experimentally, and the
superfluid-like coherence between these $p-$band atoms has been
successfully observed \cite{bloch2007}.

Orbital degeneracy also exists in materials. It is well-known that
in transition metal oxides the $d-$level orbitals are usually
partially filled, thus orbital dynamics plays an important role
(for a recent review, see reference \cite{khaliullinreview}). It
has been proposed that the quantum fluctuation of orbital
occupancy can lead to many interesting physics like ``orbital
liquids" \cite{khaliullinliquid}. However, in crystalline
materials the quantum fluctuation which tends to disorder the
orbital occupancy always competes with the Jahn-Teller effect
which distorts the lattice and favors orbital ordered state
(especially in the $e_g$ level), therefore quantum fluctuation is
only substantially strong in certain materials with $t_{2g}$ level
degeneracy (for instance, $\mathrm{LaTiO_3}$
\cite{keimer2000,khaliullinliquid}). However, in optical lattices
formed by laser beams there is no Jahn-Teller effect, thus it is
expected that new and interesting physics driven by quantum
fluctuations can emerge naturally in cold atom systems on high
orbital levels in optical lattices.

The wave function of $p_x$ state $\phi_x(r)\sim x\exp(-\alpha r^2)
$ extends further in the $x$ than in the $y$ and $z$ directions,
and so will preferentially hop to adjacent wells along the
$x-$axis. Also, $p_y$ and $p_z$ particles tend to hop in $y$ and
$z$ directions respectively. If the cubic symmetry of the lattice
is broken down to the planar square symmetry (for instance the
amplitude of laser beams propagating in the $z$ direction is
stronger than those in the $x$ and $y$ directions), the first
excited $p-$band state is two fold degenerate, with only $p_x$ and
$p_y$ levels. Also, two $p_x$ particles can interact and convert
to two $p_y$ particles, although single particle conversion is
forbidden by the reflection symmetry of the square lattice. Thus
the Hubbard type Hamiltonian describing the system can be written
as \beqn H = \sum_{i} t_a b^\dagger_{1,i}b_{1,i+\hat{x}} - t_b
b^\dagger_{1,i}b_{1,i+\hat{y}} + H.c. \cr\cr t_a
b^\dagger_{2,i}b_{2,i+\hat{y}} - t_b
b^\dagger_{2,i}b_{2,i+\hat{x}} + H.c. \cr\cr + g
b^\dagger_{1,i}b^\dagger_{1,i}b_{2,i}b_{2,i}+ H.c. + \cdots.
\label{hubbard}\eeqn Here $b_{1,i}$ and $b_{2,i}$ are annihilation
operators for $p_x$ and $p_y$ atoms on site $i$ respectively. The
ellipses above include the density-density repulsive interactions
generated from $s$-wave scattering between atoms. In this
Hamiltonian (\ref{hubbard}) $t_a
> t_b
>0$. If the boson operators are replaced by rotor operators
through $b_{x,i} \sim (-1)^{i_x}\exp(-i\theta_x)$ and $b_{y,i}
\sim (-1)^{i_y}\exp(-i\theta_x)$, the Hamiltonian becomes the
coupled anisotropic XY model \beqn H_c = \sum_i
-J_a\cos(\nabla_x\theta_{i,1}) - J_b\cos(\nabla_y\theta_{i,1})
\cr\cr -J_b\cos(\nabla_x\theta_{i,2}) -
J_a\cos(\nabla_y\theta_{i,2}) + \gamma\cos(2\theta_{i,1} -
2\theta_{i,2}). \label{coupledXYa}\eeqn In the above Hamiltonian
$J_a \sim t_a$ and $J_b \sim t_b$ are generally unequal. Besides
the classical model, we will also study the quantum version of
this model by turning on the onsite repulsion energy between boson
numbers $n_{i,\alpha}$, which is the canonical conjugate variable
of $\theta_{i,\alpha}$: \beqn H_q = H_c +
\sum_{i,\alpha}u(n_{i,\alpha} - \bar{n})^2,\label{qcoupledxya}
\eeqn $\alpha = 1,2$ denotes two orbital flavors, and $\bar{n}$ is
the average filling of each orbital flavor on each site.

The pair conversion term in (\ref{hubbard}) breaks the particle
conservation in each orbital flavor of bosons. If all the
coefficients in (\ref{coupledXYa}) are finite, the symmetry of
equation (\ref{coupledXYa}) is $U(1)\times \mathbb{Z}_2$. A
$\mathbb{Z}_2$ variable $\sigma$ can be defined as $\theta_2 =
\theta_1 + \sigma\pi/2$. The $U(1)$ phase of the system is $\chi =
\theta_1 + \theta_2$. The ground state of Hamiltonian
(\ref{coupledXYa}) is a superfluid with both $U(1)$ and
$\mathbb{Z}_2$ symmetry broken. In the rest of the paper we are
going to study the classical phase transitions between the
superfluid and disordered phase at finite temperature, as well as
the quantum phase transition between the superfluid and the Mott
Insulator by tuning the repulsive charging energy $u$.

\section{finite temperature phase transitions}

\subsection{isotropic limit: $J_a = J_b$}

In this section we study the finite temperature transition in
model (\ref{coupledXYa}). In the isotropic limit with $J_a = J_b$,
the system becomes the isotropic coupled XY model.

\beqn H =\sum_{i, \alpha,\mu} -J\cos(\nabla_\mu \theta_{i,\alpha})
+ \gamma \cos(2\theta_{1,i} - 2\theta_{2,i}).\label{coupledXY}
\eeqn

In the absence of the pair conversion term $\gamma$, at low and
finite temperature, the system has algebraic order of both
$\exp(i\theta_1)$ and $\exp(i\theta_2)$, i.e. the correlation
function
$\langle\exp(i\theta_\alpha)_{r_1},\exp(-i\theta_\alpha)_{r_2}\rangle$
falls off as a power law with distance. The coupling between the
two XY variables are always relevant in the whole algebraic phase,
which implies that no matter how weak $\gamma$ is, at large scale
$\gamma$ is always renormalized strong, and at long enough length
scale one can replace $\theta_2 = \theta_1 + \sigma \pi/2$, with
$\sigma = \pm 1$. At this length scale, the effective Hamiltonian
describing the coupled $U(1)$ and $\mathbb{Z}_2$ variables is
\beqn H_{eff} = \sum_{<i,j>} -J(1 + \sigma_i\sigma_j)\cos(\theta_i
- \theta_j). \label{isingxy}\eeqn Now $\theta = \theta_1$, and $i$
and $j$ are positions of blocks after rescaling the lattice
constant.

Many physical systems can be described by models (\ref{coupledXY})
and (\ref{isingxy}). For instance, the $J_1-J_2$ antiferromagnetic
XY model on the square lattice with nearest neighbor interaction
$J_1$ and next nearest neighbor interaction $J_2$ breaks up into
two square sublattices in its ground states, each is ordered
antiferromagnetically as long as $2 J_2 > J_1$. Effective
interaction generated by the second order spin wave perturbation
theory favors parallel aligning between spins on the two
sublattices, which breaks the $U(1)\times U(1)$ symmetry of the
classical ground states down to $U(1)\times \mathbb{Z}_2$ symmetry
\cite{henley1989,coleman1990}. Another example is the fully
frustrated XY spin model on square (or triangulr) lattice. The
classical ground states of the system break both the spin $U(1)$
symmetry and the $\mathbb{Z}_2$ symmetry of the chirality related
to the rotation of spins around each unit square
\cite{lee1984,lee1984a}. An interesting fully frustrated model
with $J_1-J_2$ interactions was also studied \cite{franzese}.

Because in the superfluid phase both $U(1)$ and $\mathbb{Z}_2$
symmetry are broken while in the high temperature disordered phase
both symmetries are restored, it is conceivable that at finite
temperature there are an Ising transition with respect to $\sigma$
and a Kosterlitz-Thouless transition for $\theta$ driven by the
proliferation of the vortices of $\theta$. A vortex of $\theta$ is
also a hybrid vortex of $\theta_1$ and $\theta_2$. A lot of
efforts have been devoted to study the sequence of the two
transitions \cite{kosterlitz1991,kosterlitz1991a}. Although it is
still premature to make a conclusion, the most recent large size
Monte-Carlo simulation suggests that there are two transitions at
finite temperature, the Kosterlitz-Thouless transition is followed
by an Ising transition at slightly higher temperature
\cite{hasenbusch2005}, with $\Delta T_c/T_c \sim 0.02$. Also, a
theoretical argument based on the fact that the cluster
percolation transition and the Ising transition coincide in the
two dimensional classical Ising model indicates that the Ising
transition should always occur at higher temperature than the
Kosterlitz-Thouless transition \cite{Korshunov2002}. It is noticed
that in Hamiltonian (\ref{isingxy}) the XY coupling across an
Ising domain wall is zero, thus the algebraic order of $
\exp(i\theta)$ can only exist in an infinite Ising domain. Because
the density of infinite Ising domain vanishes to zero at the Ising
critical point \cite{nappi1977}, the Kosterlitz-Thouless
transition can only occur in the Ising ordered phase. Thus at the
isotropic limit the Ising transition probably occurs at higher
temperature than the XY transition.

\subsection{anisotropic limit: $J_b = 0$}

In the anisotropic limit with $J_b = 0$, the ground state is
superfluid order in both $\theta_1$ and $\theta_2$ with a
subextensive $\mathbb{Z}_2$ degeneracy: one can perform
transformation $\theta_1 \rightarrow \theta_1 + \pi$ ($\theta_2
\rightarrow \theta_2 + \pi$) along any $y$ ($x$) row, the energy
of the state is unchanged \cite{girvin2005}. Let us define Ising
variable $\sigma$ as $\theta_2 = \theta_1 + \sigma \pi/2$, and
define $\theta = \theta_1$ (although the U(1) phase of the system
is $\chi = \theta_1 + \theta_2$, it is more convenient to define
$\theta = \theta_1$ to study the physics). It is noticed that in
the anisotropic limit, without the $\gamma$ term in
(\ref{coupledXYa}) both $\theta_1$ and $\theta_2$ are disordered
at finite temperature, therefore an infinitesimal $\gamma$ is not
necessarily relevant. Hereafter we assume $\gamma$ is strong
enough to guarantee that the two XY variables are perpendicular to
each other in the temperature range we are interested in, for
instance $\gamma > J_a$. The effective classical Ising-XY coupled
Hamiltonian reads \beqn H = \sum_i -J\cos(\nabla_x\theta_i) -
J\sigma_i\sigma_{i+\hat{y}}\cos(\nabla_y\theta_i).
\label{jb=0}\eeqn All the derivatives in the Hamiltonian are
defined on the lattice. This model can be viewed as the XY
variable $\theta$ coupled with an Ising gauge field background,
and the Ising gauge field background might introduce frustration
to the system. The lowest energy state can only be obtained if
there is no frustration for $\theta$, i.e. the Ising gauge flux
through each plaquette is zero. The nonfrustration criterion
generates an effective coupling between the Ising variables as
\beqn H_{ising} = \sum_{\square}
-K\sigma_1\sigma_2\sigma_3\sigma_4. \eeqn In this Hamiltonian, the
summation is over all the unit square, and $\sigma_i$, $i =
1,2,3,4$ are Ising variables defined on four corners of each unit
square. This model was first introduced in the context of $p \pm
ip$ superconductor arrays \cite{moore2004,xu2004}. The ground
state order parameter of this Ising Hamiltonian is the bond order
$b_{i,\mu } = \sigma_i\sigma_{i+\mu}$, this bond order parameter
takes nonzero expectation value along every row and column in the
whole lattice at zero temperature. However, any infinitesimal
thermal fluctuation will disorder the bond order \cite{moore2004}.

\begin{figure}
\includegraphics[width=2.5in]{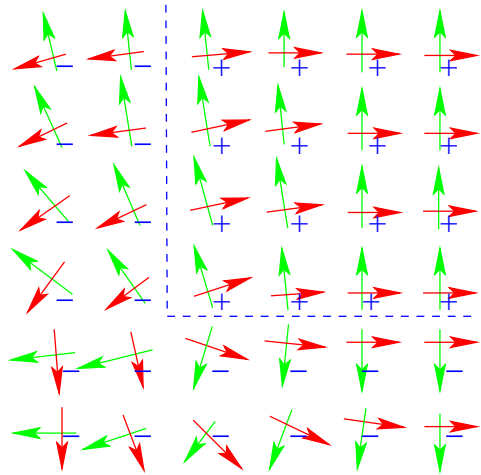}
\caption{An isolated hybrid half vortex, combined with two
infinite straight $\mathbb{Z}_2$ domain walls. The green (red)
arrows stand for $\theta_1$ ($\theta_2$).} \label{hybridvortex}
\end{figure}

The finite temperature transition is driven by the $U(1)$ defects
and the $\mathbb{Z}_2$ defects. Let us pick one ordered state from
all the subextensive degenerate states with $\sigma = 1$
everywhere. The lowest energy $U(1)$ defect has been shown to be a
hybrid half vortex \cite{girvin2005} with infinite straight
$\mathbb{Z}_2$ domain walls in both $x$ and $y$ directions without
any string tension, which is also shown in Fig.
\ref{hybridvortex}. The straight infinite $\mathbb{Z}_2$ domain
walls violates the bond order $\langle b \rangle$ along that row
and column, but as the bond order disappears at any finite
temperature, the hybrid half vortices deconfine at finite
temperature.

\begin{figure}
\includegraphics[width=2.5in]{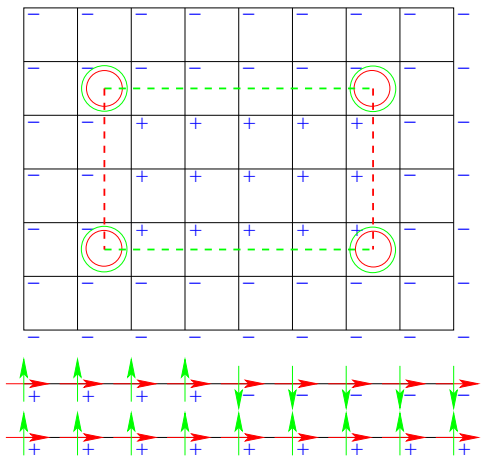}
\caption{The $\mathbb{Z}_2$ defects. The top figure is the $Z_2$
domain wall, with one hybrid half vortex attached to each corner.
The green dashed line is the $\pi$ phase mismatch of $\theta_1$,
and the red dashed line is the $\pi$ phase mismatch of $\theta_2$.
The $\mathbb{Z}_2$ domain walls will proliferate at infinitesimal
temperature. The bottom figure is the $\mathbb{Z}_2$ kink which
also proliferates under thermal fluctuation.} \label{z2domainwall}
\end{figure}

The lowest energy $\mathbb{Z}_2$ domain wall is shown in Fig.
\ref{z2domainwall}. At every corner of the $\mathbb{Z}_2$ domain
wall there is a hybrid half vortex attached. This implies that
although the $\mathbb{Z}_2$ domain wall has no string tension, the
corners are interacting with each other through a logarithmic
interaction. However, since the configuration entropy of the
domain wall is linear with its length at finite temperature, the
entropy always dominates the logarithmic interaction, thus the
domain wall will proliferate at finite temperature.

From Fig. \ref{hybridvortex} and Fig. \ref{z2domainwall}, one can
notice that across a $\mathbb{Z}_2$ domain wall along $x$
direction, $\theta_1 = \theta$ increases by $\pi$; while across a
domain wall along $y$ direction $\theta_1$ is unchanged on
average. On the contrary, $\theta_2$ gains a $\pi$-flip across a
$\mathbb{Z}_2$ domain wall along $y$ while no $\pi$-flip across
$\mathbb{Z}_2$ domain walls along $x$. Thus the U(1) phase $\chi =
\theta_1 + \theta_2$ is disordered after the proliferation of
$\mathbb{Z}_2$ domain walls, but since the superfluid stiffness is
not affected by the $\mathbb{Z}_2$ domain wall, $2\chi$ is still
algebraically ordered.

There is another type of $\mathbb{Z}_2$ defect, which is a
$\mathbb{Z}_2$ kink along one $x$ row (see Fig.
\ref{z2domainwall}). This kink defect only costs finite energy,
and will also proliferate at finite temperature due to thermal
fluctuation (very similar to the Ising kinks in one dimensional
classical Ising model), which drives the disorder of $\theta$
along $x$ axis. Thus one can conclude that at small finite
temperature $\exp(i\theta)$ has no algebraic order, while
$\exp(i2\theta)$ is algebraically ordered. The correlation length
of $\theta$ along $x$ direction is similar to that of the one
dimensional Ising model when $T \ll J$, $\xi \sim \exp(2J/T)$;
while along $y$ direction the correlation length is zero.

Thus in the anisotropic limit with $J_b = 0$, the single finite
temperature transition is a Kosterlitz-Thouless transition driven
by the proliferation of hybrid half vortices shown in Fig.
\ref{hybridvortex}. It is well-known that the proliferation of
half vortices leads to a universal stiffness jump at the
transition, which is four times as large as the stiffness jump
driven by the proliferation of single vortices: \beqn \Delta
\rho_s / T_c \sim 8/\pi. \eeqn Here the stiffness is defined as
the energy response to the same twisted boundary condition of both
$\theta_1$ and $\theta_2$. Recently half vortex proliferation has
also been predicted for the polar state in the two dimensional $s
= 1$ spinor boson condensate \cite{xu2006}.

The model (\ref{jb=0}) can be actually viewed as a ``fixed gauge"
version of the model with XY variables coupled with an Ising gauge
field. Let us consider the following Hamiltonian \beqn H_{gauge} =
\sum_{<i,j>}-J s_{i,j}\cos(\phi_i - \phi_j).\label{z2gauge} \eeqn
$s_{i,j} = \pm 1$ is an Ising variable defined on the link between
sites $i$ and $j$. It is well-known that this model has deconfined
half vortices, and the finite temperature transition is
accompanied with a universal stiffness jump $\Delta \rho_s / T_c
\sim 8/\pi$. If now we define following gauge transformations:
$\exp(i\theta_j) = \eta_j\exp(i\phi_j)$, $\eta_i = \prod_{j,j_y =
i_y, j_x < i_x}s_{j,j+\hat{x}}$, $s_{i,i+\hat{y}} =
\eta_i\eta_{i+\hat{y}}b_{i,i+\hat{y}}$, and $b_{i,i+\hat{y}} =
\sigma_i\sigma_{i+\hat{y}}$, the Hamiltonian (\ref{z2gauge})
becomes Hamiltonian (\ref{jb=0}). Here $j_x$ and $j_y$ are $x$ and
$y$ coordinates of site $j$. After the gauge transformation, an
isolated half vortex with the $\pi-$mismatch string in the $y$
direction becomes a half vortex with the string in the $x$
direction.

\subsection{$J_a > J_b > 0$, and phase diagram}

Now if a small $J_b$ is turned up, after we introduce the Ising
variable the Ising-XY Hamiltonian becomes \beqn H = \sum_i
-J(1+\beta \sigma_i\sigma_{i+\hat{x}})\cos(\theta_i -
\theta_{i+\hat{x}}) \cr\cr - J(\sigma_i\sigma_{i+\hat{y}} +
\beta)\cos(\theta_i - \theta_{i+\hat{y}}). \label{jbfinite} \eeqn
$\beta = J_b/J_a$. A small string tension proportional to $J_b$ is
added to the $\mathbb{Z}_2$ domain wall in Fig.
\ref{z2domainwall}. The domain wall energy will compete with the
configuration entropy of the domain wall, and the competition
leads to a 2D Ising transition at temperature $T \sim J_b$ which
separates an Ising ordered phase at low temperature and an Ising
disordered phase at higher temperature. A similar Ising transition
was found in another model with Hamiltonian $H = \sum_\mu-
J_b\cos(\nabla_\mu\theta) - J_a\cos(2\nabla_\mu\theta)$ and
$J_b\ll J_a$  \cite{lee1985}. the Ising transition in this model
occurs at temperature $T \sim J_b$, and the proliferation of the
Ising domain walls implies the deconfinement of half vortices.
However, in our case, since every corner of the Ising domain wall
is attached with a hybrid half vortex, the deconfinement of half
vortices would imply the proliferation of straight infinite Ising
domain walls, which cost infinite energy but the configuration
entropy is still low. Another way to look at this confinement is
that, with small $\beta$, the bond order operator $b_{i,\mu} =
\sigma_i\sigma_{i+\hat{\mu}}$ always takes nonzero expectation
value at any finite temperature, thus an infinite straight domain
wall which destroys the bond order is not favored at any
temperature. Thus even if the Ising domain wall proliferates at
high temperature, the hybrid half vortices are still confined. The
Kosterlitz-Thouless transition is still driven by the hybrid
single vortices without any infinite $\mathbb{Z}_2$ domain wall,
which leads to a universal stiffness jump $\Delta \rho_s / T_c
\sim 2/\pi$. For the same reason, the $\mathbb{Z}_2$ kink defects
are also always confined at finite temperature.

\begin{figure}
\includegraphics[width=2.8in]{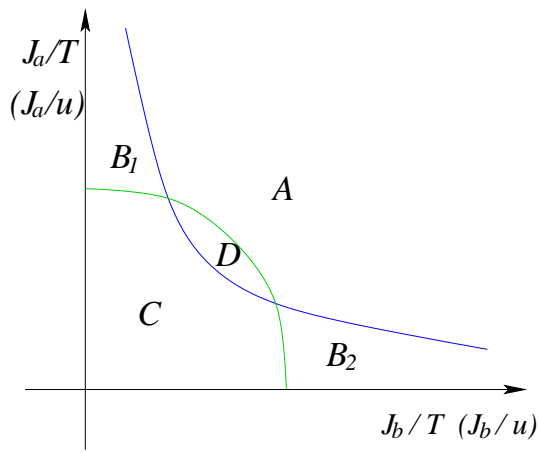}
\caption{The phase diagram for both finite temperature classical
transitions and zero temperature quantum phase transitions. The
green line is the XY transition, and the blue line is the Ising
transition. Phase $A$ is the ordered phase of both XY and Ising
variables (at finite temperature there is only algebraic order of
XY variables). Phase $C$ is the disordered phase. Phase $D$ is the
phase with Ising order, and disordered XY variables. Phase $B_1$
and $B_2$ is the Ising and $\chi$ disordered phase, but $2\chi$ is
still ordered.} \label{phasedia}
\end{figure}

It is noticed that after the Ising variables are disordered, the
XY variable $\exp(i\theta_1)$ and $\exp(i\theta_2)$ cannot be
algebraically ordered on the whole $xy$ plane. Because of the
symmetry between $\theta_1$ and $\theta_2$, the algebraic order of
$\exp(i\theta_1)$ would imply the same type of order of
$\exp(i\theta_2)$, which leads to the order (at least algebraic
order) of Ising variable $\sigma \sim i\exp(i\theta_1 -
i\theta_2)$. Also, since $\chi = \theta_1 + \theta_2$ always gains
$\pi$-flip across the Ising domain wall, the U(1) phase $\chi$ is
also disordered after the Ising transition, while $2\chi$ is still
ordered. The global finite temperature phase diagram is given by
Fig. \ref{phasedia}. When $J_b \ll J_a$, the Ising transition
occurs at temperature $T \sim J_b$, which is lower than the
Kosterlitz-Thouless XY transition. When $J_b$ is increasing, the
Ising transition finally intersects with the XY transition and the
sequence of the transitions are reversed close to the isotropic
point $J_a = J_b$. The whole phase diagram is symmetric for
interchanging $J_a$ and $J_b$.

\begin{figure}
\includegraphics[width=3.4in]{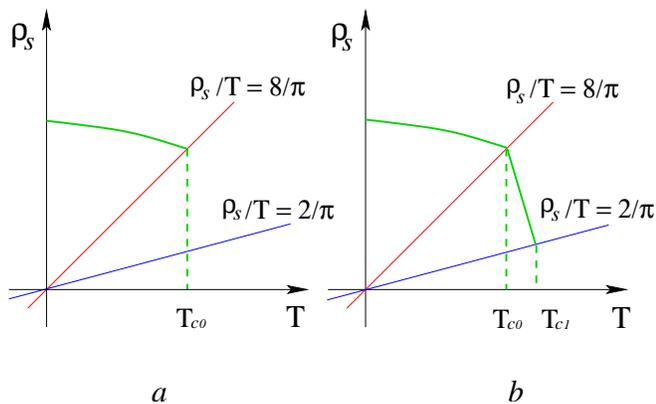}
\caption{The superfluid stiffness change at the XY transition with
zero ($a$) and nonzero but small ($b$) coupling $J_b$. When $J_b =
0$ the transition occurs at $T = T_{c0}$, the stiffness jump is
given by $\Delta \rho_s / T_c = 8/\pi$; with small $J_b$, although
the real transition occurs at $T_{c1}$, and the stiffness jump is
$\Delta \rho_s / T_c = 2/\pi$, between $T_{c0}$ and $T_{c1}$ the
change of the stiffness is steep, thus the experimental
observation is still similar to the case with $J_b = 0$. }
\label{stiffness}
\end{figure}

Although the hybrid half vortices are always confined, as long as
$J_b$ is small enough, the experimental observation should be
similar to the anisotropic limit with $J_b = 0$. For instance, let
us assume that the Kosterlitz-Thouless transition with $J_b = 0$
occurs at temperature $T_{c0}$, and the critical XY transition
temperature with small $J_b$ is $T_{c1}$, then although there is
no stiffness jump at $T_{c0}$, the slope of the stiffness change
is expected to be very steep in the window $T_{c0} < T < T_{c1}$
(Fig. \ref{stiffness}). The change of the critical temperature $
(T_{c1} - T_{c0})/T_{c0}$ can be estimated as follows: With small
$J_b$, although the hybrid half vortices are confined, the average
distance between them is still large. Let us assume the average
distance between the confined half vortices is $l_0$, which is
much greater than the lattice constant. If the correlation length
of the system is shorter than $l_0$, the system would behave as if
there were deconfined hybrid half vortices. In the absence of
$J_b$, the correlation length when $ T > T_{c0}$ scales as $\xi
\sim \exp(\sqrt{c / t})$, $t = T-T_{c0}/T_{c0}$, and $c$ is a
constant number. The scaling formula of the correlation length is
significantly changed with the presence of $J_b$ only if $\xi$
given above is larger than $l_0$, i.e. $t < c/(\ln l_0)^2$. Thus
the true Kosterlitz-Thouless transition driven by hybrid vortices
at $T_{c1}$ can only occur within this small temperature window.
Therefore $(T_{c1} - T_{c0})/T_{c0} \sim 1/(\ln l_0)^2$.

The average distance between half vortices $l_0$ can be estimated
as follows: since the half vortices are attached to the corners of
$\mathbb{Z}_2$ domain walls, let us consider four half vortices
with positive vorticity at four corners of a rectangular
$\mathbb{Z}_2$ domain wall. If the perimeter of the rectangle is
$L$, then the $\mathbb{Z}_2$ domain wall costs energy $\sim J_b
L$, while the repulsive logarithmic interaction between the half
vortices gains energy $\sim J_a \ln L$. Also, the number of
configurations for rectangles is linear with $L$, i.e. the entropy
for the rectangular $\mathbb{Z}_2$ domain wall is about $\sim \ln
L$. Thus the total free energy cost for this defect configuration
is $F \sim J_b L - (C_1 J_a + C_2 T) \ln L$, which is minimized
when $L = (C_1J_a + C_2T)/ J_b$. Here $C_1$ and $C_2$ are
nonuniversal numbers of order 1. Thus close to the XY critical
point, the average distance between half vortices is $l_0 \sim J_a
/ J_b$. Combining the results in this paragraph and last
paragraph, the change of the critical temperature due to $J_b$ and
the slope of the stiffness change between $T_{c0}$ and $T_{c1}$
are \beqn \frac{T_{c1} - T_{c0}}{T_{c0}} \sim \frac{1}{[\ln (J_a /
J_b)]^2},\cr\cr \frac{\Delta \rho_s}{\Delta T_c} \sim [\ln (J_a /
J_b)]^2. \eeqn Thus when $J_b / J_a$ is small, the stiffness
change still looks like a jump driven by half vortex
proliferation. The stiffness change and jump plotted in Fig.
\ref{stiffness} can be checked numerically \cite{xusu}.

Experimentally the ratio $J_b / J_a$ can be tuned by changing the
depth of the wells. Superfluid stiffness jump was observed in
superfluid helium \cite{heliumkt1978}, and recently the
Kosterlitz-Thouless transition has been observed in spinless
atomic Bose-Einstein condensates \cite{atomkt2006}. Our result
about the stiffness change close to the XY transition is expected
to be checked directly by future experiments.

\section{quantum phase transitions}

Now let us study the quantum phase transition at zero temperature
by tuning the ratio between the repulsion energy $u$ and the
hopping energy $J$, and we will focus on the simplest case with
integer average filling $\bar{n}$, i.e. the system has
particle-hole symmetry, more general quantum phase transitions
will be discussed in another separate work \cite{xuquantum}. The
Mott insulator phase at $\bar{n} = 1$ has been studied in
reference \cite{girvin2005}, which is a featureless Mott state
with one $p_x$ particle and one $p_y$ particle per site. In the
Mott insulator phase neither $U(1)$ nor $\mathbb{Z}_2$ symmetry is
broken. This quantum problem can be mapped to a classical problem
in 3 dimensional space. In the isotropic case with $J_a = J_b =
J$, the corresponding three dimensional model is simply model
(\ref{coupledXY}) with spatial couplings in all three directions.
In three dimensional space, although the argument based on the
percolation of clusters \cite{Korshunov2002} is no longer
applicable (because the density of infinite Ising domains is
always finite in 3D Ising model \cite{nappi1977}), the $U(1)$
transition is still expected to occur at lower temperature than
the Ising transition, since the order of $\theta_1$ and $\theta_2$
implies the order of $\sigma \sim i \exp[i(\theta_1 - \theta_2)]$.
Thus when $u/J$ is tuned from small to large, the quantum XY
transition (belonging to the 3D XY universality class) will occur
before the quantum Ising transition (belonging to the 3D Ising
universality class). However, the separation between the two
transitions is expected to be very tiny (smaller than the two
dimensional case), because a mean field Ginzburg-Landau theory for
coupled XY spin systems with the same coupling for two XY spins
always gives rise to one single transition, and the mean field
result is expected to be more accurate in three dimensional
systems than in two dimensional systems.

In the anisotropic limit with $J_b = 0$, although the Ising bond
order disappears with infinitesimal temperature, it does survive
from small quantum fluctuation. Now the effective Hamiltonian
describing the Ising variables can be written as \beqn H =
\sum_\square -K \sigma^z_1\sigma^z_2\sigma^z_3\sigma^z_4 - \sum_i
h \sigma^x_i, \label{q4spin} \eeqn and $K \sim J$, $h\sim u$. The
bond order $\langle b_{i,\mu} \rangle$ is nonzero with small
$h/K$, because infinite order of perturbation of $h/K$ is
necessary in order to mix two classical ground states. It was
conjectured that the model (\ref{q4spin}) has a quantum phase
transition at $J = h$ which separates a bond ordered phase and a
disordered phase \cite{xu2004}, and later on the nature of the
quantum phase transition was studied based on an exact mapping to
fermionic degrees of freedom \cite{chen2006}. In the bond ordered
phase, the hybrid half vortices are confined. The XY transition is
driven by the hybrid single vortex condensation if it occurs in
the bond ordered phase, otherwise it is driven by the condensation
of hybrid half vortices.

\begin{figure}
\includegraphics[width=2.8in]{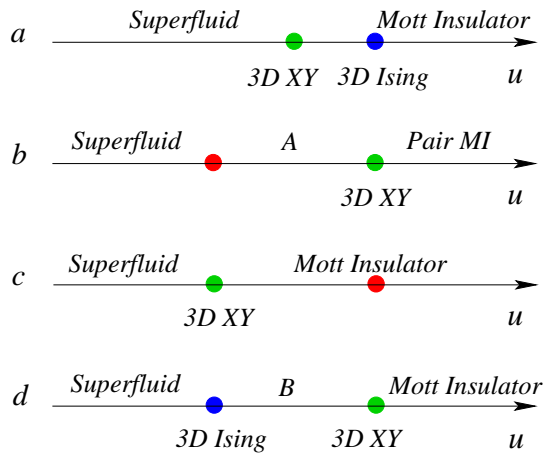}
\caption{The phase diagram at zero temperature with increasing
charging energy $u$ along $+x$ direction. $a$, $J_a = J_b = J$,
now the 3D XY transition occurs before the 3D Ising transition.
$b$, when $J_b = 0$, there are generally two transitions, one of
which is the bond order transition at zero temperature (denoted by
the red closed circle). If this transition occurs first, the
hybrid half vortices are deconfined. Phase $A$ is the bond
disordered phase and $2\chi$ ordered phase. After the condensation
of hybrid half vortices, the resultant state is the Mott Insulator
of pair bosons. $c$, when $J_b = 0$, and the bond order transition
occurs after the XY transition, the XY transition is driven by the
condensation of hybrid single vortices. $d$, the phase diagram
with small $J_b$. A quantum Ising transition occurs at $u \sim
J_b$, and the bond order parameter is always nonzero, thus the
half vortices are always confined, the XY transition is driven by
hybrid single vortex condensation. In phase $B$ after the Ising
disordering, $\chi$ is disordered while $2\chi$ is ordered.}
\label{qphasedia}
\end{figure}

If the XY transition occurs in the bond disordered phase, in the
dual formalism this transition can be described by Hamiltonian $H
= \sum_\mu -t\cos(\nabla_\mu\phi - \frac{1}{2} a_{1,\mu} -
\frac{1}{2} a_{2,\mu})$, $\phi$ is the phase angle of the
annihilation operator of hybrid half vortex, and the fluxes of
$a_{1,\mu}$ and $a_{2,\mu}$ are the boson densities $n_1$ and
$n_2$ respectively. The condensate of the vortices is the Mott
Insulator in the original boson language. One can see that after
the condensation of hybrid half vortices, an excitation in the
condensate is a vortex of $\phi$, which always involves two boson
numbers $n_1 + n_2 = \oint \vec{a}_1\cdot d\vec{l} + \oint
\vec{a}_2\cdot d\vec{l} = 2 n_v$, $n_v$ is the number of the
vortices. Thus the condensate of hybrid half vortices is a Mott
Insulator of paired bosons. The possible phase diagrams are shown
in Fig. \ref{qphasedia}.

\begin{figure}
\includegraphics[width=2.8in]{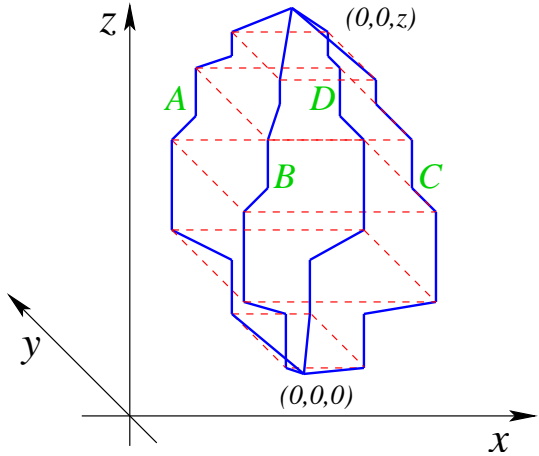}
\caption{An isolated Ising domain surrounded by world lines of
four half vortices created at $(0,0,0)$ and recombine to a single
vortex at $(0,0,z)$. The shape of the 3 dimensional Ising domain
is fully determined by the world lines $A$ and $C$. In every $xy$
plane the for hybrid-half vortices at located at four corners of a
rectangle.} \label{3dspacetime}
\end{figure}

If a small $J_b$ is turned on, a quantum Ising transition
belonging to the 3D Ising universality class is expected to occur
at $u \sim J_b$, and after the transition $\chi = \theta_1 +
\theta_2$ is disordered while $2\chi$ is ordered. It was shown in
reference \cite{balents1997} that, if an Ising system is coupled
with superfluid, the universality class of the Ising transition is
unchanged if the superfluid has particle hole symmetry, which is
satisfied in the case we are discussing. As in the finite
temperature case, with small $J_b$, hybrid half vortices are
always confined, and the XY transition is driven by hybrid single
vortices condensation. After the quantum Ising transition, the
bond order-disorder transition no longer exists, because the bond
order parameter $b_{i,\mu} = \sigma^z_i\sigma^z_{i+\mu}$ always
takes nonzero expectation values, hence the kink defect in Fig.
\ref{z2domainwall} will never proliferate. The confinement of half
vortices and $\mathbb{Z}_2$ kinks can also be understood after we
map the system to a 3D classical problem, now a deconfined hybrid
half vortex and a deconfined kink become defect lines (world lines
of half vortex and kink) along the $z$ direction, and the energy
of these defect lines in the 3D space-time always dominate the
entropy. For instance, Let us imagine a single vortex with
vorticity $+2$ be split into four half vortices at $(0,0,0)$, the
four half vortices evolve in the 3D space time and recombine into
a single vortex at $(0,0,z)$ (Fig. \ref{3dspacetime}). In any $xy$
plane between $(0,0,0)$ and $(0,0,z)$ the four half vortices are
at four corners of a rectangle. The 3 dimensional volume enclosed
by the world lines is an isolated Ising domain. The energy cost of
this Ising domain is proportional to its boundary area, while the
entropy is linear with the length of the half vortex world lines
(for instance, the shape of the whole Ising domain is fully
determined by world lines $A$ and $C$ in Fig. \ref{3dspacetime}),
i.e. the entropy is dominated by the energy and hybrid-half
vortices never deconfine. The confinement of $\mathbb{Z}_2$ kinks
can be understood similarly.

Therefore we conclude that after the quantum Ising transition both
$\mathbb{Z}_2$ variable $\sigma = i\exp(i\theta_1 - i\theta_2)$
and U(1) phase $\chi = \theta_1 + \theta_2$ are disordered, while
$2\chi$ is still ordered. The topology of the phase diagram at
zero temperature is the same as the finite temperature phase
diagram (Fig. \ref{phasedia}). The analytical results and phase
diagram Fig. \ref{qphasedia} will be checked numerically in
another work \cite{xusu}.

\section{conclusion, experiments and other physical systems}

The phases and phase transitions studied in this work can be
detected experimentally by measuring the momentum distribution
function $\langle n_k \rangle$ and measuring the momentum density
correlation $\langle n_k n_{k^\prime} \rangle$ of the
quasi-equilibrium state of $p-$band cold atoms. In real
experimental systems $J_b$ is always nonzero, let us assume $J_b$
is small and take the quantum phase transition phase diagram as an
example (case $d$ in Fig. \ref{qphasedia}). In the phase where
both $\sigma$ and $\theta$ are ordered, the momentum distribution
at small momentum is \beqn \langle n_k \rangle \sim \sum_{\vec{G}}
N k_x^2\delta^2(\vec{k} - (\pi, 0) - \vec{G}) + \cr
Nk_y^2\delta^2(\vec{k} - (0, \pi) - \vec{G}) + 2N
k_xk_y\delta^2(\vec{k} - (\pi, \pi) - \vec{G}) .
\label{mdf01}\eeqn $\vec{G}$ is the basis of the reciprocal
lattice. After the Ising transition at $u \sim J_b$, the sharp
coherence of $\theta_1$ and $\theta_2$ is erased, the best way to
characterize the system is through the momentum density
correlation function $\langle n_k n_{k^\prime} \rangle$ has also
been proposed as a tool of measuring the many body effect of cold
atom systems \cite{demler2003}, and this technique has been
successfully implemented to detect crystalline density order
\cite{folling2005}. Although $\theta_1$ and $\theta_2$ are not
ordered after the Ising transitions, $2\theta_1$ and $2\theta_2$
are still ordered. Thus sharp coherence peaks still exist in the
momentum density correlation functions \beqn\langle n_k
n_{k^\prime} \rangle \sim \langle n_k\rangle\langle
n_{k^\prime}\rangle + \delta^2(\vec{k} - \vec{k^\prime}) \langle
n_k \rangle \cr \cr + c_2 N \sum_{\vec{G}}(k_xk_x^\prime +
k_yk_y^\prime)^2\delta^2(\vec{k} + \vec{k}^\prime - \vec{G}) +
\cdots\eeqn although one has to subtract the effect of the first
two terms of this formula from the signal. The peaks in the
correlation function disappear after the XY transitions. Thus by
making use of both single particle momentum distribution function
and momentum density correlation function one can detect the phase
transitions studied in this work.

In conclusion, we studied the phase transitions of the coupled
anisotropic XY models. Both the classical and quantum phase
transitions are split into an XY transition and an Ising
transition, the global phase diagrams were drawn. The results
obtained in this work can be checked numerically and hopefully by
future experiments on $p-$band cold atom systems. Throughout the
paper we assumed that the cubic symmetry is broken down to the
square lattice symmetry and only $p_x$ and $p_y$ orbital levels
are considered. If the optical lattice has perfect cubic symmetry,
then all three $p-$wave states should be taken into account.
However, the pair conversion term between all three $p-$wave
states $\gamma \{ \cos[2(\theta_1 - \theta_2)] + \cos[2(\theta_2 -
\theta_3)] + \cos[2(\theta_3 - \theta_1)]\}$ is frustrating when
$\gamma > 0$, the three terms cannot be minimized simultaneously,
which makes the problem more complicated.

Similar problems can be studied for atoms on even higher orbital
levels. For instance, in cubic lattice the $d-$level state on each
well is split into a three fold degenerate $t_{2g}$ level and a
two fold degenerate $e_g$ level. Within the $e_g$ level, the
$d_{x^2 - y^2}$ state extends further and isotropically in the
$xy$ plane, while the $d_{3z^2 - r^2}$ level extends in the $z$
direction. Thus in the isotropic limit the problem becomes the
coupled isotropic XY model in three dimensional space, and as
discussed in section III, the 3D Ising transition is expected to
occur at higher temperature than the 3D XY transition. On the
contrary, in the anisotropic limit with extreme directional
hoppings, the finite temperature model can be written as \beqn H =
\sum_i\sum_{\mu = x,y} -J\cos(\nabla_\mu \theta_{i,1}) - J
\cos(\nabla_z\theta_{i,2}) \cr + \gamma \cos(2\theta_1 -
2\theta_2). \eeqn If the Ising variable is introduced through
$\theta_2 = \theta_1 + \sigma \pi/2$, the Ising-XY model becomes
\beqn H = \sum_i\sum_{\mu = x,y} -J\cos(\nabla_\mu \theta_i) - J
\sigma_i\sigma_{i+\hat{z}}\cos(\nabla_z\theta_i). \eeqn Again the
nonfrustration criterion generates an effective classical Ising
Hamiltonian \beqn H = \sum_{i} - J
\sigma_{i}\sigma_{i+\hat{x}}\sigma_{i+\hat{z}}\sigma_{i+\hat{x}+\hat{z}}
\cr\cr - J
\sigma_{i}\sigma_{i+\hat{y}}\sigma_{i+\hat{z}}\sigma_{i+\hat{y}+\hat{z}}.
\label{egising} \eeqn If we define bond operator $b_{i} =
\sigma_i\sigma_{i+z}$, the model (\ref{egising}) becomes decoupled
2D classical Ising model for new Ising variable $b_i$. Thus in the
anisotropic limit there are again two transitions at finite
temperature, one is the 3D XY transition, and the other is a 2D
Ising transition.

\begin{center}
\bf{Acknowledgement}
\end{center}

The author thanks Joel E. Moore, Subroto Mukerjee and Daniel
Podolsky for helpful discussions. This work is supported by the
National Science Foundation through NSF grant DMR-0238760.

\bibliography{ptransition03}
\end{document}